\documentstyle[aps,prl]{revtex}
\begin{document}


\wideabs{

\title{The s--wave implosion and explosion of free
particles}
\author{I. Bia\l ynicki--Birula$^{1,2}$, M. A. Cirone $^{2}$,
J. P. Dahl$^{2,3}$, M. Fedorov$^{2,4}$,
W. P. Schleich$^{2}$}
\address{$^1$ Center for Theoretical Physics,
Polish Academy of Sciences, Al. Lotnik\'ow
32/46, 02-668, Warsaw, Poland \\
$^2$ Abteilung f\"ur Quantenphysik,
Universit\"at Ulm, D-89069 Ulm, Germany \\
$^3$ Chemical Physics, Department of Chemistry,
Technical University of Denmark, DTU 207, DK-2800
Lyngby, Denmark \\
$^4$ General Physics Institute, Russian Academy
of Sciences, 38 Vavilov st., Moscow 117942 Russia}
\date{\today}
\maketitle

\begin{abstract}
A free quantum particle in two dimensions with vanishing
angular momentum (s--wave) in the form
of a ring--shaped wave packet feels
an attractive force towards the center of the ring, leading
first to an implosion followed by an explosion.
\end{abstract}

\vspace{0.5cm}
}

The spreading of a wave packet is an important topic in
quantum mechanics \cite{bohm}. It arises
due to different momenta contained in the
packet. As pointed out by Max Born \cite{born} even a classical
ensemble of particles spreads.
The only quantum effect lies in the connection between
the position and the momentum distributions, that is,
the spreads in position and momentum
are not independent.
In the present paper we predict a wave effect of the
free particle that manifests itself in the dynamics:
A ring--shaped wave packet in two dimensions of vanishing
angular momentum feels an attractive force towards the
center of the ring. This effect is unique to two dimensions
and manifests itself most strongly for vanishing angular momentum.

In order to illustrate this surprising phenomenon we
first consider the free motion of wave packets in $N$
dimensions. We then concentrate on two and three dimensions, where
we propagate a circular shell and a spherical shell, respectively.
All wave packets have initially vanishing angular momentum
and vanishing average radial momentum. We show that the circular
shell first implodes and then explodes. In contrast,
the spherical shell does not implode but only explodes.
We demonstrate this wave effect unique to two--dimensional
space using ({\em i}) analytical expressions for the average radial
position and momentum,
({\em ii}) the Wigner function approach
\cite{wign} towards quantum mechanics,
({\em iii}) analytical expressions \cite{us2} for the radial
Green's function in two dimensions and ({\em iv})
the numerical propagation of wave packets.

Our analytical approach considers the propagation of
the initial wave function

\begin{equation}
\Phi^{(N)}_{0}(x_{1},\ldots,x_{N})=\frac{{\cal N}}{\sqrt{S^{(N)}}}
r^{\gamma}
\exp \left[-\frac{1}{2}\left(\frac{r}{\delta r}\right)^2
\right]
\label{eq1}
\end{equation}
in the absence of any potential. Here ${\cal N}\equiv \left[
2/\Gamma\left( N/2+\gamma \right)\right]^{1/2}\delta r^{-N/2-\gamma}$
denotes the radial normalization constant and
$r\equiv \left(x_{1}^{2}+x_{2}^{2}+ \ldots x_{N}^{2}\right)^{1/2}$
is the radial variable in $N$ space dimensions.
Moreover, $S^{(N)}\equiv 2\pi^{N/2}/\Gamma\left( N/2 \right)$
is the total solid angle.
The quantity $\delta r$ defines a characteristic length scale of
the packet and is a measure of the radial width \cite{rem1}.

We first take the exponent $\gamma$ to be equal to 2.
This special form of the initial wave function -- polynomial
in coordinates multiplied by a Gaussian -- allows us
to calculate analytically the time dependence of the average
radial position $\langle r^{(N)} \rangle$ and the average radial
momentum $\langle p^{(N)} \rangle$ in $N$ dimensions.
We then turn to the general case of arbitrary $\gamma>1$
but concentrate on two dimensions.

In the derivation of our results we follow three independent approaches that provide
identical results: ({\em i}) We consider the time evolution in Wigner phase space and
calculate the averages with the help of the Wigner function; ({\em ii}) we use the method
of the generating function \cite{bia2} to propagate a modulated Gaussian wave packet of
the free particle in $N$ dimensions, and evaluate the averages; ({\em iii}) we propagate
the free wave packet with the help of the radial Green's function of the free particle in
two dimensions. Here we do not present the detailed calculations but only summarize the
results \cite{us3}.

We start by presenting first the general results for the propagation
in $N$ dimensions and then focus on the special case of two
and three dimensions.
It is convenient to introduce the dimensionless time variable
$\tau \equiv \hbar t/(\delta r^2 M)$, where $M$ denotes the
mass of the particle. We find \cite{us3} the expression

\begin{equation}
\langle r^{(N)} \rangle (\tau) =
\frac{1+\tau^2/a^{(N)}}
{\sqrt{1+\tau^2}} r_{0}^{(N)}
\label{iwo}
\end{equation}
for the average radial position. Here we have introduced
the abbreviation

\begin{equation}
a^{(N)}\equiv 1+\frac{4N}{N^2+3}
\label{an}
\end{equation}
and the initial radial position

\begin{equation}
r_{0}^{(N)}\equiv \frac{\Gamma
\left(\frac{N+5}{2}\right)} {\Gamma\left( \frac{N+4}{2}\right)}
\delta r,
\end{equation}
in $N$ dimensions, where $\Gamma$ denotes the Euler gamma function.

Equation (\ref{iwo}) shows that for large times
the average radial position increases linearly with time,
corresponding to an explosion of the wave packet.
This holds true for any number of dimensions.

For short times we can expand the square root
in Eq. (\ref{iwo}) which yields

\begin{equation}
\langle r^{(N)} \rangle (\tau) \simeq
\left[1+\left( \frac{1}{a^{(N)}}-\frac{1}{2}\right) \tau^2\right]
\, r^{(N)}_{0}.
\end{equation}
An interesting situation occurs when $a^{(N)}>2$. In this case,
the average radius initially decreases, corresponding
to an implosion of the wave packet. From the definition Eq. (\ref{an})
of $a^{(N)}$ we arrive at the implosion condition

\begin{equation}
(N-1)(N-3)<0.
\label{ineq}
\end{equation}
Hence an implosion
occurs only for $N=2$ corresponding to $a^{(2)}=15/7=2+1/7$.

The phenomenon of the implosion can also be viewed by
considering the time evolution of the average radial momentum

\begin{equation}
\langle p^{(N)} \rangle (\tau) = M \frac{d}{dt} \langle r^{(N)} \rangle
=\frac{d}{d\tau} \langle r^{(N)} \rangle \frac{\hbar}{\delta r^2}.
\end{equation}
determined by the time derivative of the average position
$\langle r^{(N)} \rangle$, which yields

\begin{equation}
\langle p^{(N)} \rangle (\tau)=-
\frac{\tau}{(1+\tau^2)^{3/2}} \left(
a^{(N)}-2-\tau^2 \right)\, p_{\infty}^{(N)}
\label{mome}
\end{equation}
where

\begin{equation}
p_{\infty}^{(N)} \equiv \lim_{\tau \rightarrow \infty} p^{(N)}(\tau)
=\frac{1}{a^{(N)}}\frac{\hbar r_{0}^{(N)}}{\delta r^2 }.
\end{equation}

The radial momentum is negative, corresponding to an implosion,
provided $\tau < \left[a^{(N)}-2\right]^{1/2}$.
Since in two dimensions $a^{(2)}=15/7$ this condition reads
$\tau < \tau_{\rm min} \equiv 1/\sqrt{7}$. At that moment
the momentum vanishes and the implosion turns into an explosion.
The average radial position assumes a minimum $r_{\rm min}\equiv
\langle r^{(2)}\rangle \left(\tau_{\rm min}\right)$.

In order to highlight the special property of two dimensions
we now compare the dynamics of the wave packet Eq. (\ref{eq1})
for $\gamma=2$ in two and three dimensions.
In Fig. \ref{fig1} we show the time evolution of the average
radial position and the average radial momentum
in two (solid curves) and three (dashed curves) dimensions
based on Eqs. (\ref{iwo}) and (\ref{mome})
with $a^{(2)}=15/7$ and $a^{(3)}=2$.

For short times the average radius of a ring--shaped
wave packet decreases as a function of
time. This initial implosion of the circular wave is
followed by a final explosion.
We emphasize that this phenomenon is not due to an external
classical force but occurs for a free particle moving
in two dimensions \cite{rem2}.
Obviously the spreading of the ring--shaped wave packet in
the radial direction is asymmetric; initially the probability flux
towards the center is larger than towards the outside.

In the three--dimensional case, where we propagate a spherical shell,
the average position and the average momentum do not display
the soft implosion effect of two dimensions.
Indeed, since $a^{(3)}=2$, the average momentum
is always positive and
the average position cannot decrease but must increase.

The contraction effect is maximal at $\tau_{\rm
min}=1/\sqrt{7}$ and according to Eq. (\ref{iwo})
the average radial position
$\langle r^{(2)} \rangle$ assumes the minimal value

\begin{equation}
r_{\rm min}\equiv \sqrt{\frac{224}{225}}r_{0}^{(2)}\simeq
 0.9978 \, r_{0}^{(2)}.
\end{equation}

Hence, it is a rather soft implosion.

We note that this result is independent of $\delta r$,
since there is only one length
scale in the problem. When we choose a different form for
the packet we can make the effect larger.
For example, by replacing $r^\gamma$ in Eq. (\ref{eq1}) by
 $\sin(r^2/\delta r^2)$, we obtain at $\tau_{\rm min}\simeq 1.11$

\begin{equation}
r_{\rm min} \simeq 0.9964\,r_{0}^{(2)}.
\end{equation}

In order to study the implosion effect we concentrate
on two dimensions and now consider an arbitrary value
of the exponent $\gamma>1$ in the the wave packet
Eq. (\ref{eq1}). With the help of the corresponding
radial propagator \cite{us2} we obtain the radial
wave function which yields the expression

\begin{equation}
\langle r^{(2)} \rangle (\tau) = 2^\gamma
\frac{\Gamma^2\left( \frac{\gamma+1}{2} \right)}{\Gamma\left(
\gamma+\frac{1}{2} \right)}
\frac{\tau^{\gamma-1}}{\left(1+\tau^2 \right)^{\frac{\gamma+1}{2}}}
I(\tau;\gamma) r_{0}^{(2)}
\end{equation}
for the average radial position. Here the integral

\begin{eqnarray}
I(\tau;\gamma) & \equiv & \int_{0}^{\infty} \!\!\!\!\! d \xi \, \xi^2
\exp\left( -\frac{1}{1+\tau^2}\, \xi^2 \right) \nonumber \\
& & \left| _{1} F_{1}\left[ -\frac{\gamma-1}{2};1;
\frac{i}{2\tau}\frac{1}{1+i\tau}
\left(\frac{r}{\delta r}\right)^2 \right] \right|^2
\end{eqnarray}
containing the confluent hypergeometric function
$_{1}F_{1}$
can be expressed in terms of the generalized
hypergeometric function $_{3} F_{2}$. However, we found it more
convenient to evaluate this integral numerically.
For each value of $\gamma$ we obtain an implosion but the ratio
$r_{\rm min}/r^{(2)}_{0}$ is always of the same order of magnitude.

We now turn to the explanation of the
implosion effect using the Wigner function \cite{wign}.
The free particle is one of the few quantum mechanical
systems in which the classical phase space dynamics
is identical to the quantum mechanical one.
Indeed, the Wigner function satisfies
the classical Liouville equation. Hence, we find
the Wigner function at time $t$ from the Wigner function
at time $t=0$ by replacing the position $\vec{r}$
by $\vec{r}-\vec{p}\,t/M$. We emphasize that now $\vec{r}$ and
$\vec{p}$ denote $N$--dimensional vectors. The corresponding
probability distributions in position and momentum follow
by integration over the conjugate variable.

Since for the free particle classical and quantum phase
space dynamics are identical the
quantum effect of implosion rather than explosion is
stored in the initial Wigner
function. To be more precise, it is in the correlations
between position and momentum. A
classical phase space distribution enjoys a probability
interpretation and therefore
always has to be positive. In contrast, the Wigner
function can assume negative values.
Indeed, both the Wigner functions of the circular shell
and of the spherical shell
contain domains in phase space where they become
negative. Since the Wigner function is
normalized the total volume of the Wigner function is unity.
We can calculate numerically
the volumes $V^{(2)}_{-}$ and $V^{(3)}_{-}$ corresponding
to the negative parts. We find
in the two dimensional case $V^{(2)}_{-}\simeq 0.27$,
whereas in three dimension
$V^{(3)}_{-}\simeq 0.23$. Hence, there is a slightly
larger contribution of negative parts
in two dimensions than in three dimensions.

In the phase space description of quantum mechanics
quantum effects
are stored in the negative parts of the initial Wigner
function. Since a classical ensemble of particles cannot have
a negative phase space distribution it cannot display the
implosion effect.
Needless to say a classical ensemble with a ring--shaped
initial position distribution can show an implosion
provided there exists a non--vanishing negative initial
radial momentum. However, in our quantum mechanical wave
packet the average radial momentum vanishes.

For the choice Eq. (\ref{eq1}) of the initial
wave packet the radial probability density

\begin{eqnarray}
W^{(N)}(r)dr & = & \mid \Phi^{(N)}_{0}(r)\mid^2 r^{N-1}dr S^{(N)}
\nonumber \\
& = & \mid {\cal N} \mid^2 r^{N+2\gamma-1}e^{-(r/\delta r)^2} dr
\end{eqnarray}
depends on the number of dimensions.
Hence, one might argue that the implosion is a
result of the slightly different initial radial wave
functions. In order to exclude this argument we have
performed a numerical integration of the Schr\"odinger equation

\begin{equation}
i\hbar \frac{\partial}{\partial t}u^{(N)}_{0}(r,t)=
\left[ -\frac{\hbar^2}{2M}\frac{\partial^2}{\partial r^2}
+V^{(N)}(r)\right] u^{(N)}_{0}(r,t)
\end{equation}
for the radial wave function $u^{(N)}_{0}$ of the free particle
in two and three space dimensions, corresponding to zero angular momentum.
Here we always start with the same initial radial wave function.
In this way we guarantee identical radial probabilities densities
in two and three dimensions.

The radial potential
\begin{equation}
V^{(N)}(r)\equiv \frac{\hbar^2}{2M}
\frac{(N-1)(N-3)}{4r^2}
\label{vndim}
\end{equation}
in $N$ dimensions involves the product $(N-1)(N-3)$
familiar from the implosion condition Eq. (\ref{ineq}).

In three dimensions the potential $V^{(3)}$ vanishes.
However, for $N=2$ the  quantum anti--centrifugal
potential

\begin{equation}
V_{Q}(r)\equiv V^{(2)}(r)\equiv -\frac{\hbar^2}{2M}\frac{1}{4r^2}
\label{quan}
\end{equation}
does not vanish and is negative, that is, attractive \cite{flug,us}.
This potential is responsible for the implosion.

A large class of wave packets with vanishing angular
momentum displays this effect of first imploding
and then exploding \cite{comm}. Indeed, in the
insert of Fig. \ref{fig1} we show the short time evolution
of the average radial position and the average
radial momentum in two and three dimensions for an initial
radial Gaussian wave packet $u(r,t=0)\equiv \tilde{\cal N}
\exp \left[ -\left( r-\rho \right)^2/\delta r^2 \right]$
displaced by an amount $\rho$ from the origin.
As in the analytical results we find
an implosion of a circular but not of a spherical shell.

So far we have concentrated on a single particle
but the effect of implosion manifests itself also in the
relative motion of two free particles. We can envision
the total wave function of the two particles to be
of the form Eq.(\ref{eq1}) corresponding to an
entangled state. When we assume that each particle
is allowed to move in a $D$--dimensional space the total
wave function lives in a $N=2D$ dimensional
configuration space. The physical situation of a single
particle imploding in two dimensions corresponds to two
particles constrained to one dimension. When they move along a line
their average relative separation first decreases and
then increases. It looks like the particles first attract
and then repel each other. However, when they move in two
or three dimensions, their relative distance always increases.

We conclude by briefly outlining an experiment to
demonstrate the implosion effect. At first sight, one might
think of cold atoms around a wire familiar from
atom chips \cite{hind}. However, it might be easier to make
use of the analogy between the Schr\"odinger equation
and the paraxial wave equation of classical electrodynamics.
We shine monochromatic light onto an opaque screen
with an annular aperture. In order to avoid diffraction
from sharp edges, we introduce apodization, that
is, a position dependence transmission coefficient.
At the edges of the ring almost no light is transmitted,
whereas at the center of the ring almost all light passes
through.
This arrangement creates an intensity distribution after the screen
confined mainly to a ring. The subsequent free propagation
in space is analogous to the free propagation in time of the
Schr\"odinger wave function. The time dependent
probability distribution then corresponds to the
transverse intensity distributions at different locations
downstream from the screen \cite{foot}.

We thank I. Sh. Averbukh, M. V. Berry, D. Bouw\-meester, H. Carmichael,
D. Greenberger, D. Kobe, R. F. O'Connell, W. C. Schieve,  F. Straub,
S. Varro and K. W\'odkiewicz for many fruitful discussions.
I. B.--B., J. P. D. and M. F.
gratefully acknowledge the support of the Humboldt foundation
and the great hospitality enjoyed at the
Abteilung f\"ur Quantenphysik.
The work of W. P. S. is partially supported by DFG.

\begin{figure}
\caption{Comparison between the time evolutions of a
circular (solid line) and a spherical shell
(dashed line)
using the average radial position (top) and the average radial
momentum (bottom) obtained from Eqs.(\ref{iwo})
and (\ref{mome}) for $a^{(2)}=15/7$ and $a^{(3)}=2$.
On average the two--dimensional shell
first contracts and then expands, whereas the
three--dimensional one expands from the onset.
In the left inserts we show the short time
behavior of a displaced Gaussian radial wave packet
obtained numerically.
Here we have used the same Gaussian for an initial radial
wave function in two and three dimensions with
$\delta r=0.4$ and $\rho=1.5$ in arbitrary units.
Position is scaled in units of the initial
position $r_{0}^{(N)}$ of the
wave packet, and momentum in terms of the final
momentum $p_{\infty}^{(N)}$.}
\label{fig1}
\end{figure}



\begin{references}
\bibitem{comm} A purely Gaussian wave packet
located initially at the origin does not implode. It spreads symmetrically in radial
direction. From Heisenberg's equations of motion for the radial variable or from
Ehrenfest's theorem we can derive conditions on the wave function for the implosion
effect to occur, see I. Bia\l ynicki--Birula, J. P. Dahl, and W. P. Schleich, to be
published.
\bibitem{foot} The implosion effect
is not the wave mechanics analogue
of the Poisson spot \cite{bowo} familiar from the diffraction
theory of classical optics. The implosion results from the
modification factor ${\cal H}$ discussed in Ref.\cite{us2}
in Eq.(15) of the free Green's function.
A Poisson spot occurs already when we approximate
the radial propagator in two dimensions
by the difference of two one--dimensional Green's functions
of free motion. Hence, the implosion effect is a modification
on top of the Poisson spot.
\end{references}
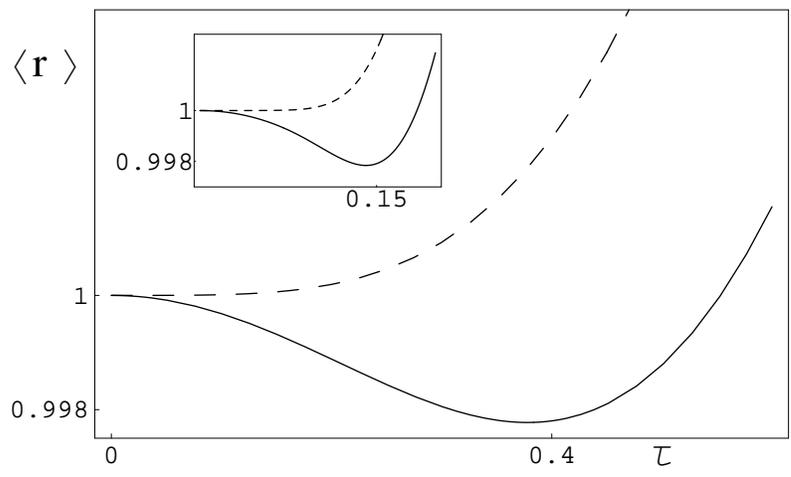
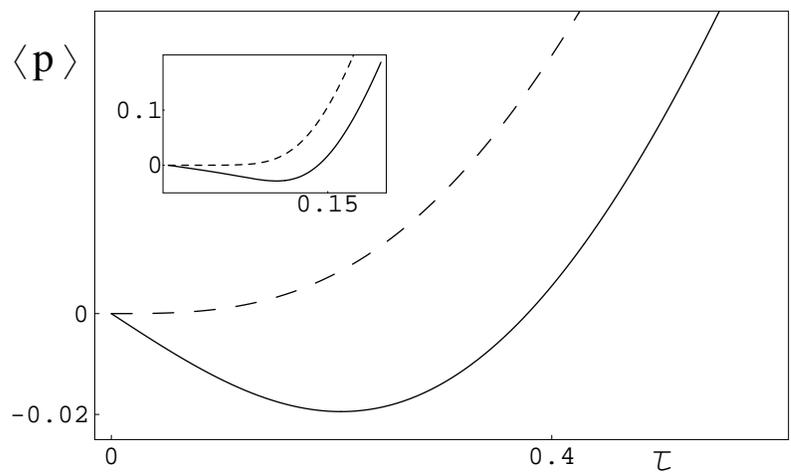
\end{document}